\documentclass[preprint,12pt]{elsarticle}

\usepackage{natbib}

\usepackage{graphics}
\usepackage{epsfig}

\usepackage{apj-abbr}

\def\sax{{\em BeppoSAX\/}}

\usepackage{amssymb}

\begin{document}

\begin{frontmatter}



\title{Spectral evolution of GRBs observed with \sax\ WFCs and GRBM}


\author{F. Frontera}
\address{University of Ferrara, Physics Department, Via Saragat, 1 44100 Ferrara, Italy
and INAF, IASF, Via Gobetti, 101, 40129 Bologna, Italy}
\ead{frontera@fe.infn.it}
%
\author{L. Amati}
\address{INAF, IASF, Via Gobetti, 101, 40129 Bologna, Italy}
%
\author{C. Guidorzi}
\address{University of Ferrara, Physics Department, Via Saragat, 1 44100 Ferrara, Italy}
%
\author{R. Landi}
\address{INAF, IASF, Via Gobetti, 101, 40129 Bologna, Italy}
%
\author{V. La Parola}
\address{INAF, IASF, Via La Malfa, 153, 90146 Palermo, Italy}

\begin{abstract}
We present some preliminary results obtained from a systematic analysis of almost all
GRBs simultaneously observed with the Gamma Ray Burst Monitor and the
Wide Field Cameras aboard the \sax\ satellite. 

\end{abstract}


\end{frontmatter}

\parindent=0.5 cm

\section{Motivations}

In spite of the huge advances in the knowledge ot the GRB afterglow properties, 
the GRB phenomenon is still poorly understood. Of crucial importance, it is recognized 
to be the study of the prompt emission, which is directly connected with 
the original explosion. 
The radiation emission mechanism at work is still matter of debate. 
Most best fit models are still empirical: power-law (PL), smoothly broken PL proposed by
\citet{Band93}(Band law, BL), power--law with high energy exponential cutoff (CPL). 
Physical models have also been proposed (e.g., synchrotron shock model \citep{Tavani96}, BB$+$PL).

Relations between intrinsic peak energy $E_{p,i}$ and GRB released energy/luminosity 
(e.g., Amati relation \cite{Amati02}  or the Yonetoku relation \cite{Yonetoku04}) 
could help to solve the radiation mechanism issue.  
However these relations, derived using the GRB time integrated spectral properties,  
are still debated, mainly due to the dispersion of the data points
around the best fit power--law. It is well known that the spectral properties evolve with time. 
What happens about these relations when the time resolved properties are considered?

Recently several authors have concentrated their interest on the time resolved spectra, 
using mainly BATSE data \citep{Ghirlanda07,Ryde08,Peng09b}.
According to Ryde \& Pe'er \citep{Ryde08}, the time resolved spectra of GRB FRED-like 
pulses can be fit with a BB$+$PL. However the BATSE data cover the hard X-/soft gamma-ray
energy band (20-2000 keV). What happens if the band is extended  down to 2 keV? 

We present here some preliminary results of the  time resolved spectral analysis of the entire 
sample of GRBs observed with BeppoSAX WFC plus GRBM in the 2-700 keV band. 
Results on the time resolved properties of 8 GRBs were published by \citet{Frontera00}.

\section{The GRB sample}

\begin{figure}
  \includegraphics[height=.3\textheight]{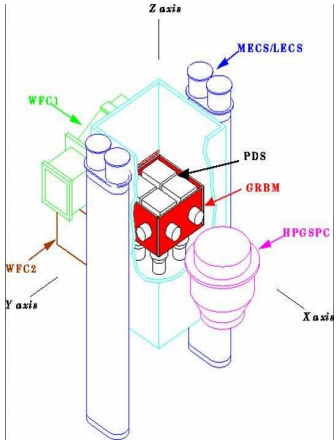}
  \caption{The \sax\ payload in which the location of the GRBM and WFCs is shown.}
\label{f:payload}
\end{figure}

The entire GRB sample is made of 55 events simultaneously detected with both
the Gamma Ray Burst Monitor (GRBM) and the Wide Field Cameras
(WFCs) aboard the BeppoSAX satellite. 
An exhausive description of the GRBM can be found in \citet{Frontera09a} and references therein,
while that of the WFCs can be found in \citet{Jager97}. 
The main features of the GRBM 
were the following: an energy band from  40 to 700 keV, a Field of View (FOV) of about 2$\pi$ sr, 
a time resolution, in the case of a GRB trigger, down to 0.5 ms in the 40-700 keV
energy channel, and a continuous transmission of the counts integrated over 1 s in 2 
energy channels (40-700 keV, >100 keV) and over 128 s in 240 energy channels to cover the 40--700 keV
band.
Concerning the WFCs, their main features were the following: an energy band from 2 to 28 keV, a 
FOV of 40x40 deg (FWZR), 31 energy channels with a time resolution 
of 0.5 ms. A view of the \sax\ payload is shown in Fig.~\ref{f:payload}.

%
%

\begin{figure}
  \includegraphics[angle=270,width=.5\textwidth]{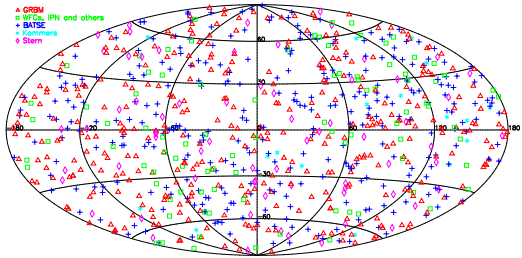}
  \caption{Sky distribution of the GRBs detected with the \sax\ GRBM. It includes those also
detected with WFCs. Reprinted from \citet{Frontera09a}.}
\label{f:catalog}
\end{figure}

The GRBM/WFC events were extracted from the population of 1082 GRBs  detected with the GRBM 
(see GRBM catalog in the paper by \citet{Frontera09a} and shown in Fig.~\ref{f:catalog}).
The 2--700 keV time-resolved spectral analysis performed thus far concerns 45 GRBs.

%
%

\section{Data analysis and fit models}

Each GRB time profile is subdivided in time slices with duration that takes into account 
the GRB pattern and the statistics of the data. Background is properly subtracted.

Spectra are derived in each of the time slices. In the spectral fits, GRBM and WFC 
normalization factors were left free to vary in the range 0.8-1.3, found in 
extended cross-calibrations. 

Many input models were tested. We found that a photoelectrically  {\sc cpl} (as called in
the XSPEC spectral deconvolution software \cite{Arnaud96}) gives the  best fits and the 
best constrained parameters of the derived spectra.
Thus we adpoted this model. However  a photoelectrically absorbed {\sc bl} was used 
when  the peak energy $E_p$ of the $E F(E)$ spectrum
derived from {\sc cpl} is inconsistent with that derived from {\sc bl} and, 
at the same time, {\sc bl} gives a better fit and a constrained value of the high energy
index $\beta$.  

\section{Preliminary results}

\subsection{Test of the blackbody plus power--law model}

\begin{figure*}
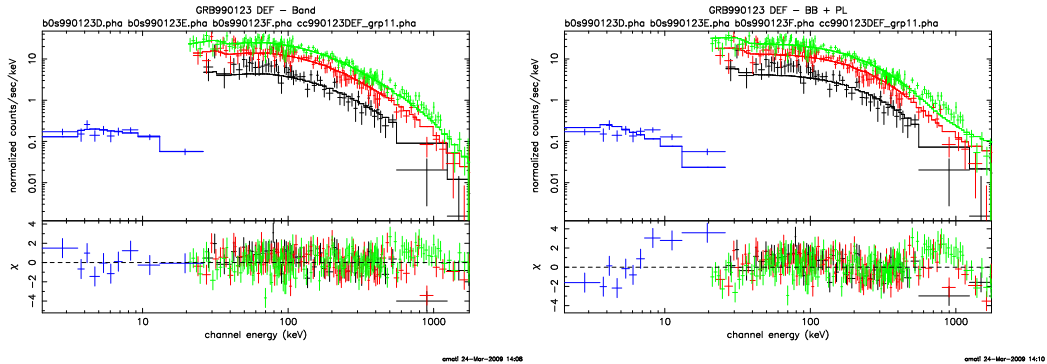

 \includegraphics[angle=270,width=.5\textwidth]{fig3a.ps}
 \includegraphics[angle=270,width=.5\textwidth]{fig3b.ps}
 \caption{Count spectrum of 6 s duration taken during the rise of GRB990123. {\em Left panel}: Fit 
with a BL model. {\em Right panel}: Fit with a BB$+$PL.}
\label{f:ryde}
\end{figure*}

We have also tested the blackbody plus power-law (BB$+$PL) model recently proposed by 
\citet{Ryde08} for the fit of the time resolved spectra of 56 strong BATSE GRBs.
If we use only BATSE data, we confirm their results for GRB\,990123), the only GRB 
of their sample observed with \sax GRBM plus WFC. However this model  is rejected when we fit  
the BATSE$+$WFC or WFC$+$GRBM time resolved spectra.
In Fig.~\ref{f:ryde} we show the results obtained in the case of the 6 s duration count
spectrum measured during the rise of GRB990123.

%
%

\subsection{Evolution of the spectral parameters with time}

\begin{figure}
 \includegraphics[height=.3\textheight]{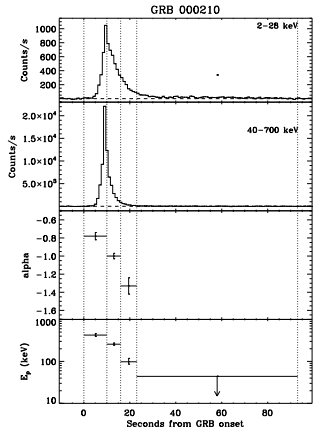}
  \caption{Light curve in two energy bands and behavior of the spectral parameters 
with time in the case of the bright event GRB\,000210.}
\label{f:000210}
\end{figure}

Depending on GRB and its brightness, the time resolved spectral parameters have 
a different behavior with time. As far as the peak energy $E_p$ of the $EF(E)$ spectrum
is concerned, its time behaviour is twofold: in some cases it mimics the GRB pattern
of the prompt emission, in other cases it decreases with time 
(e.g., Fig~\ref{f:000210}). As far as the 
low energy photon index $\alpha$ is concerned, in general it decreases with time (i.e.,
the spectrum softens), but we also found some cases in which it mimics the GRB pattern.
In any case, we do not find any correlation between $\alpha$ and $E_p$, with $\alpha$
mainly determined by the WFCs data.

%
%

\subsection{Time resolved peak energy versus flux}

\begin{figure}
 \includegraphics[height=.3\textheight]{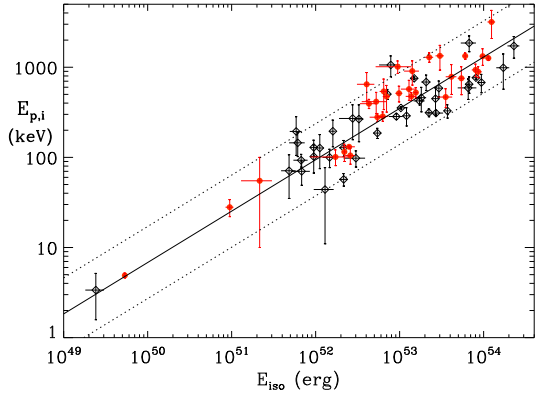}
  \caption{The $E_{p,i}$ vs. $E_{iso}$ correlation based on the time averaged spectra of a 
sample of 70 GRBs with known redshift. Reprinted from \citet{Amati08}.}
\label{f:epeiso}
\end{figure}

As it is  well known, a key importance relation is that found by us in 2002 \cite{Amati02}(now
known as Amati relation) between
intrinsic time averaged peak energy $E_{p,i}$ and time averaged relased energy $E_{iso}$ 
of the GRB prompt emission, derived assuming  isotropy.
This relation is now widely confirmed by all GRBs (about 100) with known redshift $z$ 
\cite{Amati09}, detected thus far, except the nearest GRB ever observed 
(\sax\ GRB\,980425, with $z= 0.0085$). A similar relation was later found by 
\citet{Yonetoku04}, using the peak luminosity $L_{p,iso}$, instead of the total released energy, 
evaluated from the time averaged spectrum of the prompt emission. 
Either the Amati relation or the Yonetoku relation are very robust, but they are
characterized by a significant spread, that is inconsistent with the statistical uncertainty 
in the data points (see Fig.~\ref{f:epeiso} for $E_p$ vs. $E_{iso}$). Also due to this spread, 
the Amati relation is questioned by 
some authors \cite{Band05,Butler07,Butler09,Shahmoradi09}, who state that it could be 
the result of selection effects.  However other 
authors \cite{Ghirlanda05,Ghirlanda08,Nava08,Nava09} find 
that these effects do not invalidate the relation. The time resolved spectra can help 
to settle this issue.

\begin{figure}
 \includegraphics[height=.3\textheight]{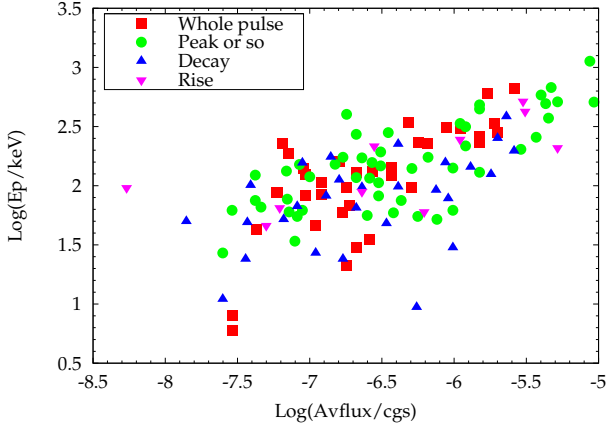}
  \caption{Time resolved  peak energy vs. 2--700 keV flux. Different colours
and simbols are used for different subinterval types within each GRB.}
\label{f:ep-flux}
\end{figure}

We find that, within each burst, the measured peak energy is related with the 2--700 keV flux 
and this dependence is clearly independent of the redshift or other selction effects. 
The $E_p$ vs. 2--700 keV flux ($F_{2-700}$) is shown 
in Fig.~\ref{f:ep-flux} for those GRBs in our sample for which constrained values of 
$E_p$ are possible. As it can be seen,
a positive correlation between $E_p$ and $F_{2-700}$  is outstanding and it is
 a clear consequence of the correlation we have found between these two quantities 
within each GRB. The best fit is obtained with power--law ({\sc pl}) 
$E_p \propto F_{2-700}^\alpha$, with index $\alpha = 0.43\pm 0.07$ and a significant 
extrinsic scatter ($\sigma_{logE_p} = 0.30$). This scatter is apparent in the figure above,
especially for fluxes lower than $10^{-6}$~cgs.
A higher contribution to the scatter seems to be due to the parameters derived during
the event tail. 

By limiting the analysis to the GRBs with known $z$, we derived, in the $z$ corrected plane,
the intrinsic time resolved peak energies $E_{p,i}$ versus the corresponding bolometric (in the
1--10000 keV rest frame energy band) luminosities $L_{bol}$. As also expected,
we find that the {\sc pl} correlation between $E_{p,i}$ and $L_{bol}$ is confirmed. The derived
best fit {\sc pl} index is consistent with 0.5, and the scattering of the
($E_p$, $F_{2-700}$) data points is almost unchanged. 

Exhaustive results of our comparative analysis are the subject of a paper in preparation 
that will be published soon (Frontera et al. 2009, in preparation).

%
%
%

%
%

\section{Conclusions}

The 2-700 keV time resolved spectra of 55 GRBs detected with BSAX WFC+GRBM are being analyzed
and compared. 

A photoelectric absorbed power--law with a high energy exponential  cutoff (CPL) appears 
to be the best model for the comparison of the time resolved spectra.
A smoothly broken power--law (Band law) is used in a few controversial cases.

A blackbody $+$ power--law model is inconsistent with the WFC data. The results found by
\citet{Ryde08} and their consequences should be revisited.

No correlation between measure peak energy $E_p$ and photon index of the {\sc cpl} is found.
This result strenghtens the $E_p$ vs. Flux correlation.

A strong power--law correlation between the measured $E_p$ and the 2-700 keV flux is 
observed, with a power--law index$ = 0.43\pm0.07$ and significant extrinsic scatter 
($\sigma_{\log E_p} = 0.30$).

A power--law index consistent with 0.5 and a similar scatter is found in the 
case of $E_{p,i}$ vs. $L_{iso}$.
Similar results, with different slope, are found when $E_{p,i}$ is correlated  with 
the GRB beaming corrected $L_\gamma$. However we find a slightly lower scatter 
($\sigma_{\log E_p} = 0.21$) in the assumption of a jet emission in a standard ISM environment. 

All these results strenghten the Amati correlation with a possible explanation 
of the spread of the data points around the best fit PL in terms of the spread of the time resolved
$E_p$ vs. Flux correlation within each GRB.

The final analysis is in progress and will be published soon.

\section{Acknowledgments}
 We wish to thank Jean J. M. in't Zand from SRON, Utrecht for supplying us the time resolved 
spectra of WFCs.
The \sax\ mission was an effort of the Italian Space Agency ASI, with
a participation of the Netherland Space Agency NIVR.

\bibliographystyle{elsarticle-harv}
\bibliography{grb_ref}

\hyphenation{Post-Script Sprin-ger}
\begin{thebibliography}{21}
\expandafter\ifx\csname natexlab\endcsname\relax\def\natexlab#1{#1}\fi
\expandafter\ifx\csname url\endcsname\relax
  \def\url#1{\texttt{#1}}\fi
\expandafter\ifx\csname urlprefix\endcsname\relax\def\urlprefix{URL }\fi

\bibitem[{{Amati} et~al.(2009){Amati}, {Frontera}, and {Guidorzi}}]{Amati09}
{Amati}, L., {Frontera}, F., {Guidorzi}, C., Jul. 2009. {Spectrum-energy
  correlations in Gamma-Ray Bursts confront extremely energetic Fermi GRBs}.
  ArXiv e-prints.

\bibitem[{{Amati} et~al.(2002){Amati}, {Frontera}, {Tavani}, {in't Zand},
  {Antonelli}, {Costa}, {Feroci}, {Guidorzi}, {Heise}, {Masetti}, {Montanari},
  {Nicastro}, {Palazzi}, {Pian}, {Piro}, and {Soffitta}}]{Amati02}
{Amati}, L., {Frontera}, F., {Tavani}, M., {in't Zand}, J.~J.~M., {Antonelli},
  A., {Costa}, E., {Feroci}, M., {Guidorzi}, C., {Heise}, J., {Masetti}, N.,
  {Montanari}, E., {Nicastro}, L., {Palazzi}, E., {Pian}, E., {Piro}, L.,
  {Soffitta}, P., Jul. 2002. {Intrinsic spectra and energetics of BeppoSAX
  Gamma-Ray Bursts with known redshifts}. \aap 390, 81--89.

\bibitem[{{Amati} et~al.(2008){Amati}, {Guidorzi}, {Frontera}, {Della Valle},
  {Finelli}, {Landi}, and {Montanari}}]{Amati08}
{Amati}, L., {Guidorzi}, C., {Frontera}, F., {Della Valle}, M., {Finelli}, F.,
  {Landi}, R., {Montanari}, E., Dec. 2008. {Measuring the cosmological
  parameters with the $E_{p,i}-E_{iso}$ correlation of gamma-ray bursts}.
  \mnras 391, 577--584.

\bibitem[{{Arnaud}(1996)}]{Arnaud96}
{Arnaud}, K.~A., 1996. {XSPEC: The First Ten Years}. In: {Jacoby}, G.~H.,
  {Barnes}, J. (Eds.), Astronomical Data Analysis Software and Systems V. Vol.
  101 of Astronomical Society of the Pacific Conference Series. pp. 17--+.

\bibitem[{{Band} et~al.(1993){Band}, {Matteson}, {Ford}, {Schaefer}, {Palmer},
  {Teegarden}, {Cline}, {Briggs}, {Paciesas}, {Pendleton}, {Fishman},
  {Kouveliotou}, {Meegan}, {Wilson}, and {Lestrade}}]{Band93}
{Band}, D., {Matteson}, J., {Ford}, L., {Schaefer}, B., {Palmer}, D.,
  {Teegarden}, B., {Cline}, T., {Briggs}, M., {Paciesas}, W., {Pendleton}, G.,
  {Fishman}, G., {Kouveliotou}, C., {Meegan}, C., {Wilson}, R., {Lestrade}, P.,
  1993. {BATSE observations of gamma-ray burst spectra. I - Spectral
  diversity}. \apj 413, 281--292.

\bibitem[{{Band} and {Preece}(2005)}]{Band05}
{Band}, D.~L., {Preece}, R.~D., Jul. 2005. {Testing the Gamma-Ray Burst Energy
  Relationships}. \apj 627, 319--323.

\bibitem[{{Butler} et~al.(2009){Butler}, {Kocevski}, and {Bloom}}]{Butler09}
{Butler}, N.~R., {Kocevski}, D., {Bloom}, J.~S., Mar. 2009. {Generalized Tests
  for Selection Effects in Gamma-Ray Burst High-Energy Correlations}. \apj 694,
  76--83.

\bibitem[{{Butler} et~al.(2007){Butler}, {Kocevski}, {Bloom}, and
  {Curtis}}]{Butler07}
{Butler}, N.~R., {Kocevski}, D., {Bloom}, J.~S., {Curtis}, J.~L., Dec. 2007. {A
  Complete Catalog of Swift Gamma-Ray Burst Spectra and Durations: Demise of a
  Physical Origin for Pre-Swift High-Energy Correlations}. \apj 671, 656--677.

\bibitem[{{Frontera} et~al.(2000){Frontera}, {Amati}, {Costa}, {Muller},
  {Pian}, {Piro}, {Soffitta}, {Tavani}, {Castro-Tirado}, {Dal Fiume}, {Feroci},
  {Heise}, {Masetti}, {Nicastro}, {Orlandini}, {Palazzi}, and
  {Sari}}]{Frontera00}
{Frontera}, F., {Amati}, L., {Costa}, E., {Muller}, J.~M., {Pian}, E., {Piro},
  L., {Soffitta}, P., {Tavani}, M., {Castro-Tirado}, A., {Dal Fiume}, D.,
  {Feroci}, M., {Heise}, J., {Masetti}, N., {Nicastro}, L., {Orlandini}, M.,
  {Palazzi}, E., {Sari}, R., Mar. 2000. {Prompt and Delayed Emission Properties
  of Gamma-Ray Bursts Observed with BeppoSAX}. \apjs 127, 59--78.

\bibitem[{{Frontera} et~al.(2009){Frontera}, {Guidorzi}, {Montanari}, {Rossi},
  {Costa}, {Feroci}, {Calura}, {Rapisarda}, {Amati}, {Carturan}, {Cinti}, {Dal
  Fiume}, {Nicastro}, and {Orlandini}}]{Frontera09a}
{Frontera}, F., {Guidorzi}, C., {Montanari}, E., {Rossi}, F., {Costa}, E.,
  {Feroci}, M., {Calura}, F., {Rapisarda}, M., {Amati}, L., {Carturan}, D.,
  {Cinti}, M.~R., {Dal Fiume}, D., {Nicastro}, L., {Orlandini}, M., Jan. 2009.
  {The Gamma--Ray Burst catalog obtained with the Gamma Ray Burst Monitor
  aboard BeppoSAX}. \apjs 180, 192--223.

\bibitem[{{Ghirlanda} et~al.(2007){Ghirlanda}, {Bosnjak}, {Ghisellini},
  {Tavecchio}, and {Firmani}}]{Ghirlanda07}
{Ghirlanda}, G., {Bosnjak}, Z., {Ghisellini}, G., {Tavecchio}, F., {Firmani},
  C., Jul. 2007. {Blackbody components in gamma-ray bursts spectra?} \mnras
  379, 73--85.

\bibitem[{{Ghirlanda} et~al.(2005){Ghirlanda}, {Ghisellini}, and
  {Firmani}}]{Ghirlanda05}
{Ghirlanda}, G., {Ghisellini}, G., {Firmani}, C., Jul. 2005. {Probing the
  existence of the $E_{peak}-E_{iso}$ correlation in long gamma ray bursts}.
  \mnras 361, L10--L14.

\bibitem[{{Ghirlanda} et~al.(2008){Ghirlanda}, {Nava}, {Ghisellini}, {Firmani},
  and {Cabrera}}]{Ghirlanda08}
{Ghirlanda}, G., {Nava}, L., {Ghisellini}, G., {Firmani}, C., {Cabrera}, J.~I.,
  Jun. 2008. {The $E_{peak}-E_{iso}$ plane of long gamma-ray bursts and
  selection effects}. \mnras 387, 319--330.

\bibitem[{{Jager} et~al.(1997){Jager}, {Mels}, {Brinkman}, {Galama},
  {Goulooze}, {Heise}, {Lowes}, {Muller}, {Naber}, {Rook}, {Schuurhof},
  {Schuurmans}, and {Wiersma}}]{Jager97}
{Jager}, R., {Mels}, W.~A., {Brinkman}, A.~C., {Galama}, M.~Y., {Goulooze}, H.,
  {Heise}, J., {Lowes}, P., {Muller}, J.~M., {Naber}, A., {Rook}, A.,
  {Schuurhof}, R., {Schuurmans}, J.~J., {Wiersma}, G., 1997. {The Wide Field
  Cameras onboard the BeppoSAX X-ray Astronomy Satellite}. \aaps 125, 557--572.

\bibitem[{{Nava} et~al.(2009){Nava}, {Ghirlanda}, and {Ghisellini}}]{Nava09}
{Nava}, L., {Ghirlanda}, G., {Ghisellini}, G., May 2009. {Selection effects on
  GRB spectral-energy correlations}. In: {Meegan}, C., {Kouveliotou}, C.,
  {Gehrels}, N. (Eds.), American Institute of Physics Conference Series. Vol.
  1133 of American Institute of Physics Conference Series. pp. 350--355.

\bibitem[{{Nava} et~al.(2008){Nava}, {Ghirlanda}, {Ghisellini}, and
  {Firmani}}]{Nava08}
{Nava}, L., {Ghirlanda}, G., {Ghisellini}, G., {Firmani}, C., Dec. 2008. {Peak
  energy of the prompt emission of long gamma-ray bursts versus their fluence
  and peak flux}. \mnras 391, 639--652.

\bibitem[{{Peng} et~al.(2009){Peng}, {Ma}, {Zhao}, {Yin}, {Fang}, and
  {Bao}}]{Peng09b}
{Peng}, Z.~Y., {Ma}, L., {Zhao}, X.~H., {Yin}, Y., {Fang}, L.~M., {Bao}, Y.~Y.,
  Jun. 2009. {The $E_{p}$ Evolutionary Slope Within the Decay Phase of ''Fast
  Rise and Exponential Decay'' Gamma-Ray Burst Pulses}. \apj 698, 417--427.

\bibitem[{{Ryde} and {Pe'er}(2008)}]{Ryde08}
{Ryde}, F., {Pe'er}, A., Nov. 2008. {Quasi-blackbody component and radiative
  efficiency of the prompt emission of gamma-ray bursts}. ArXiv e-prints.

\bibitem[{{Shahmoradi} and {Nemiroff}(2009)}]{Shahmoradi09}
{Shahmoradi}, A., {Nemiroff}, R.~J., Apr. 2009. {The Possible Impact of GRB
  Detector Thresholds on Cosmological Standard Candles}. ArXiv e-prints.

\bibitem[{{Tavani}(1996)}]{Tavani96}
{Tavani}, M., Aug. 1996. {A Shock Emission Model for Gamma-Ray Bursts. II.
  Spectral Properties}. \apj 466, 768--778.

\bibitem[{{Yonetoku} et~al.(2004){Yonetoku}, {Murakami}, {Nakamura},
  {Yamazaki}, {Inoue}, and {Ioka}}]{Yonetoku04}
{Yonetoku}, D., {Murakami}, T., {Nakamura}, T., {Yamazaki}, R., {Inoue}, A.~K.,
  {Ioka}, K., 2004. {Gamma-Ray Burst Formation Rate Inferred from the Spectral
  Peak Energy-Peak Luminosity Relation}. \apj 609, 935--951.

\end{thebibliography}

\end{document}